\documentclass[namedreferences]{SolarPhysics}
\usepackage[optionalrh]{spr-sola-addons} % For Solar Physics
\usepackage{graphicx}        % For eps figures, newer & more powerfull
\usepackage{color}           % For color text: \color command
\usepackage{url}             % For breaking URLs easily trough lines
\newcommand{\aap}{    {\it Astron. Astrophys.}}

\newcommand{\apj}{    {\it Astrophys. J.}}

\newcommand{\jgr}{    {\it J. Geophys. Res.}}

\newcommand{\pasj}{   {\it Pub. Astron. Soc. Japan}}

\newcommand{\solphys}{{\it Solar Phys.}}

\begin{document}

\begin{article}
\begin{opening}

\title{Solar Intranetwork Magnetic Elements: bipolar flux appearance}

\author{Jingxiu~\surname{Wang}$^{1}$\sep
        Guiping~\surname{Zhou}$^{1}$\sep
        Chunlan~\surname{Jin}$^{1}$\sep
        Hui~\surname{Li}$^{2}$\sep}
\runningauthor{Wang et al.} \runningtitle{Solar intranetwork
magnetic bipole appearance}

\institute{$^{1}$ Key Laboratory of Solar Activity\\
National Astronomical Observatories, Chinese Academy of Sciences,
Beijing 100012, China (email:\url{wangjx@nao.cas.cn})\\
$^{2}$ Key Laboratory of Dark Matter and Space Astronomy\\
Purple Mountain Observatory, Chinese Academy of Science,
Nanjing 210008, China}

\begin{abstract}

The current study aims to quantify characteristic features of
bipolar flux appearance of solar intranetwork (IN) magnetic
elements. To attack such a problem, we use the Narrow-band Filter
Imager (NFI) magnetograms from the Solar Optical Telescope (SOT) on
board \emph{Hinode}; these data are from quiet and an enhanced
network areas. Cluster emergence of mixed polarities and IN
ephemeral regions (ERs) are the most conspicuous forms of bipolar
flux appearance within the network. Each of the clusters is
characterized by a few well-developed ERs that are partially or
fully co-aligned in magnetic axis orientation. On average, the
sampled IN ERs have total maximum unsigned flux of several $10^{17}$
Mx, separation of $3-4$ arcsec, and a lifetime of $10-15$ minutes.
The smallest IN ERs have a maximum unsigned flux of several
$10^{16}$ Mx, separations less than 1 arcsec, and lifetimes as short
as 5 minutes. Most IN ERs exhibit a rotation of their magnetic axis
of more than 10 degrees during flux emergence. Peculiar flux
appearance, {\it e.g.}, bipole shrinkage followed by growth or the
reverse, is not unusual. A few examples show repeated
shrinkage-growth or growth-shrinkage, like magnetic floats in the
dynamic photosphere. The observed bipolar behavior seems to carry
rich information on magneto-convection in the sub-photospheric
layer.

\end{abstract}

\end{opening}

\section{Introduction}

The solar surface is often divided in active regions and quiet Sun.
The term `quiet Sun' in general refers to regions far from active
regions (ARs) where the violent activity, {\it e.g.}, flares, takes
place. However, the `quiet Sun' is never quiet. Many types of
small-scale magnetic activity have been revealed, such as
microflares (Lin et al., 1987), explosive events in the transition
region (Dere et al., 1991), X-ray bright points (Vaiana et al.,
1973), X-ray jets (Shibata et al., 1992) and mini-filament eruptions
(Wang et al., 2000). They are indicators of an exceedingly dynamic
sea of mixed-polarity magnetic fields everywhere on the Sun (Wang et
al., 2000), possibly powered by the quiet Sun non-potential magnetic
field (Zhao et al., 2009; Yang et al., 2011). Small-scale activity
on the Sun, {\it e.g.}, jets and spicules, can provide a ubiquitous
mass supply that is crucial to coronal heating (see De Pontieu et
al., 2011).

Network (NT) magnetic elements were known since the 60s (Sheeley,
1967). Intra-network or inner-network (IN) magnetic fields were
first described by Livingston and Harvey (1975) and Smithson (1975)
as `discrete elements' of mixed polarities `interior to the
network'. By mid-90s, a few papers (Keller et al., 1994; Lin, 1995;
Wang et al., 1995; Lites et al., 1996) largely renewed the interest
in the intranetwork field. Great progress has been made (see the
recent reviews by de Wijn et al., 2009 and S\'{a}nchez Almeida and
Mart\'{\i}nez Gonz\'{a}lez, 2011) since then. With the new {\it
Hinode} Solar Optical Telescope (SOT) observations, it appears
timely to re-examine the key results in this working area and to
initiate new efforts to more thoroughly understand this important
component of solar magnetism. A few key issues on the physics of the
solar intranetwork magnetic field remain to be settled down and
require further studies, such as:

\begin{enumerate}

\item Intrinsic properties, {\it e.g.}, the field strength, filling
factor, and internal structure. Are the IN elements strong or weak
in term of the equipartition magnetic field that has a rough
equipartition with the kinetic energy of photospheric plasma flow
(Lin, 1995; Solanki et al., 1996; Lin and Rimmele, 1999; S\'{a}nchez
Almeida and Lites, 2000; Khomenko et al., 2003, 2005; S\'{a}nchez
Almeida et al., 2003; Socas-Navarro and Lites, 2004; Mart\'{\i}nez
Gonz\'{a}lez et al., 2006; Orozco Su\'{a}rez et al., 2007a, b; Jin,
Wang, and Xie, 2011)? Do they still have internal magnetic structure
at the current resolution (see S\'{a}nchez Almeida and Lites, 2000)?

\item Flux appearance and disappearance. What is the nature of
horizontal magnetic elements (Lites et al., 1996; Harvey et al.,
2007; Lites et al., 2008; Ishikawa et al., 2008; Ishikawa and
Tsuneta, 2009; Jin et al., 2009a; Ishikawa, Tsuneta, and
Jur$\check{c}\acute{a}$k, 2010)? Is there a dominant flux of
intranetwork elements in their flux distribution (Wang et al.,
1995)? How important are these tiny magnetic elements for the Sun's
magnetic energy supply (Zirin, 1987; S\'{a}nchez Almeida, 1998;
Trujillo Bueno, Shchukina, and Asensio Ramos, 2004; Thornton and
Parnell, 2011; Jin et al., 2011)? How much flux can still be hidden
below the spatial scales that can be resolved (S\'{a}nchez Almeida
and Lites, 2000; S\'{a}nchez Almeida, Emonet, and Cattaneo, 2003;
Trujillo Bueno, Shchukina, and Asensio Ramos, 2004)? In which ways
does the flux disappear from the solar surface? Does flux disappear
by magnetic reconnection in the lower solar atmosphere, or by
submergence below the photosphere, or by diffusion into unresolved
scraps (Livi, Wang, and Martin, 1985; Wang and Shi, 1993; Zhang et
al., 1998a; Zhou et al., 2010)?

\item Observed characteristics of magneto-convection. Can we
envision how magneto-convection works in the sub-photospheric
convective plasma using observations (see Zhang, Yang, and Jin,
2009)? How does the local turbulent dynamo operate in the
near-surface layers of the Sun (see Cattaneo, 1999; S\'{a}nchez
Almeida, Emonet, and Cattaneo, 2003; V\"{o}gler et al., 2005; Stein
and Nordlund, 2006; V\"{o}gler and Sch\"{u}sler, 2007; Stein et al.,
2011)? Is there evidence of the convective collapse in the quiet Sun
magnetism (Solanki et al., 1996; Nagata et al., 2008)?

\item Response of the solar atmosphere to the magnetic evolution
in the quiet Sun. What is the magnetic nature of network bright
points (Dunn and Zirker, 1973; Mehltretter, 1974; Muller and
Roudier, 1984; Berger et al., 1995, 2004)? Are we confident about
the vision of ``magnetic bright points" (see S\'{a}nchez Almeida et
al., 2004; S\'{a}nchez Almeida et al., 2010)? What magnetic
quantities and processes determine the heating of the upper
atmosphere (Berger and Title, 2001; Ishikawa et al., 2007; Zhao et
al., 2009; De Pontieu et al., 2011)?
\end{enumerate}

All quantitative aspects of the above issues are critical in
constraining the models of the origin and role of the IN field in
solar magnetism and atmospheric heating.

In earlier papers, we have described the life-story and lifetime of
IN magnetic elements (Zhou et al., 2010) and studied the intrinsic
properties of intranetwork and network elements including the field
strength, filling factor, inclination, and other aspects (Jin, Wang,
and Xie, 2011). The IN elements are found to be intrinsically weak
in term of the equipartition field strength. This paper is focused
on the IN flux appearance in the form of bipolar emergence,
shrinkage and submergence. Our main motivation is to explore the
interaction of magnetic and convective fields in the photosphere and
immediate sub-photosphere.

Based on the SOT/Spectro-polarimeter (SP) and SOT/FGIV observations
onboard {\it Hinode}, which have improved the spatial resolution and
polarization sensitivity, flux appearance at IN scale has been amply
studied. Centeno et al. (2007) reported an event of bipolar
emergence within a quiet granule. Extended examples of bipolar
emergence have been studied by Mart\'{\i}nez Gonz\'{a}lez and Bellot
Rubio (2009). These authors find that a significant fraction of the
magnetic flux in IN regions appears in the form of $\Omega$-shaped
loops with an emergence rate of 1.1$\times 10^{12}$ Mx s$^{-1}$
arcsec$^{-2}$, which corresponds to 8.38$\times 10^{26}$ Mx per day
for the whole Sun. Unipolar flux emergence has been studied by Lamb
et al.(2010). They find it likely that the coalescence of weak flux
leads to the appearance of unipolar flux elements. Orozco Su\'{a}rez
et al. (2008) describe a particular example of vertical field
emergence; the emergence of horizontal magnetic elements has been
reported by many authors (see Ishikawa and Tsuneta, 2009). Ishikawa
and Tsuneta (2009) distinguished the buoyancy-driven emergence (type
1) and the convection-driven emergence (type 2). The later is
related to the horizontal magnetic elements. Zhang, Yang and Jin
(2009) examined the interaction of flux emergence and granulation.
Jin, Wang and Zhao (2009b) obtained the vector magnetic field
distribution in solar granules. These observations provide a rich
set of information on the magneto-convection in the layer
immediately beneath the photosphere. With the unprecedented spatial
resolution and polarization sensitivity from {\it Hinode} and new
ground-based observations, we seem to be at a critical moment to
diagnose or even to image the subsurface magneto-convection. It is
also worth to mention that remarkable progress in numerical
simulations of magneto-convection has been made (see Cheung et al.,
2008). The simulations have shed new light in understanding the
recent observations.

In this study we use SOT/Narrow-band Filter Imager (NFI)
magnetograms to particularly explore magnetic bipolar appearance in
IN regions. Emergence in clusters of mixed polarity and IN ephemeral
regions are exemplified, and a peculiar pattern of bipolar
shrinkage-growth, and/or growth-shrinkage, like a magnetic float in
convective plasma, is described for the first time.

The next section is devoted to the description of observations and
calibration methods. Section 3 is devoted to the detailed
description of magnetic bipolar appearance in the form of clusters,
IN ephemeral regions and float-like bipoles. Conclusions and
discussion will be presented in the last section.

\section{{\it Hinode} SOT/NFI Observations}
\subsection{Why are SOT/NFI observations needed?}

This series of studies is mostly based on {\it Hinode} SOT/NFI data.
The basic reason to choose SOT/NFI data comes from the following
considerations. SOT/NFI data have adequate temporal resolution, {\it
e.g.}, 1-2 minutes, in addition to high spatial resolution of 0.3
arcsec. We can follow many magnetic elements in a large field of
view. The selection of IN bipoles is much less ambiguous when the
temporal behavior can be taken into account. Moreover, the evolution
of the magnetic bipoles can be followed for a reasonable time
interval in order to understand their appearance and disappearance.
Using the simultaneous filtergrams obtained by {\it Hinode}/SOT, we
can diagnose the chromospheric response to the magnetic evolution, a
study that we defer for a later time.

\subsection{Selected quiet regions}

Two regions are selected for this study. They are listed in Table 1.
The pixel size for the observations of both regions is 0.16 arcsec,
and the spatial resolution is 0.32 arcsec. The analysis is made for
the time-sequence of the quiet-network region (QNT) on 24 June 2007
(see Figure 1) and the enhanced-network region (ENT) on 11 December
2006 (see Figure 2). The ENT is to the east of NOAA active region
(AR) 10930 and covers partially the AR plage. To the west in the
center of Figure 2, the outer boundary of a sunspot penumbra is
marked by dashed curved line. For the quiet network region, the
temporal resolution is 1 minute; while for the enhanced network
region it is 2 minutes. In Table 1, we list two other quiet-network
regions in lines 2 and 3 for a future study on the flux distribution
(Zhou, Wang, and Jin, 2011). All the quiet regions are located close
to disk center.

\begin{table}
\caption{The studied ENT region and QNT regions. } \label{tbl-0}
\begin{tabular}{ccccccc}
  \hline
dates  & Obs. (UT) & Wavelenth (\AA) & Pos. &  FOV (pix$^{2}$)  & Charac. \\
  \hline
Dec. 11 2006 & 19:08-23:58 &  FeI 6302 & S10E01 & 638$\times$788
& ENT \\
Dec. 19 2006 &  23:56   &  FeI 6302 & S00E10 & 626$\times$782
& QNT \\
Dec. 20 2006 &  06:36   &  FeI 6302 & S00E02 & 751$\times$783
& QNT \\
June 24 2007 & 17:12-18:09 &  NaI 5896 & S00W10 & 932$\times$894
& QNT \\

\hline
\end{tabular}
\end{table}

\begin{figure}
\centerline{\includegraphics[width=0.9\textwidth,clip=]{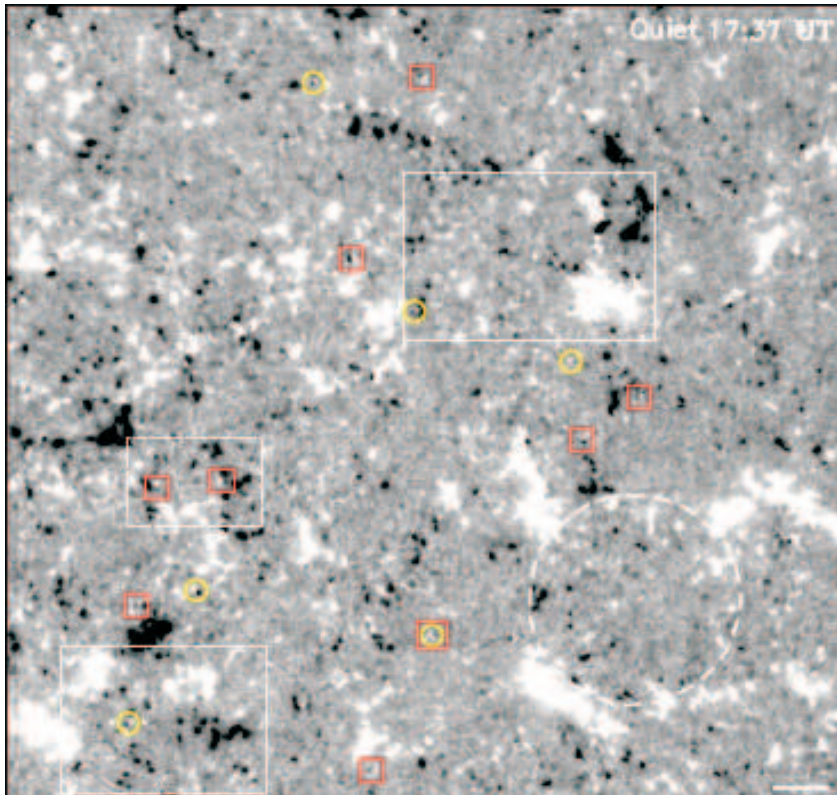}}
\caption{The SOT/NFI line-of-sight magnetogram at 17:36:50 UT, 24
June 2007, is scaled in the range of $\pm$ 50 G. A bar in the
lower-right corner indicates a 10 arcsec length. The framed boxes
correspond to the windows discussed in text; from top to bottom: IN
ERs (framed by solid lines), peculiar flux appearance (by
dashed-dotted lines), and cluster emergence of mixed polarities (by
dashed lines). In the lower part of the magnetogram, a complete
network cell is marked by a circle. A few tiny IN ERs are marked by
small yellow circles, while cancelling magnetic features (CMFs) are
identified as red squares. Here and in the following figures, north
is to the top and east, to the left.}\label{fig1}
\end{figure}

\begin{figure}
\centerline{\includegraphics[width=0.8\textwidth,clip=]{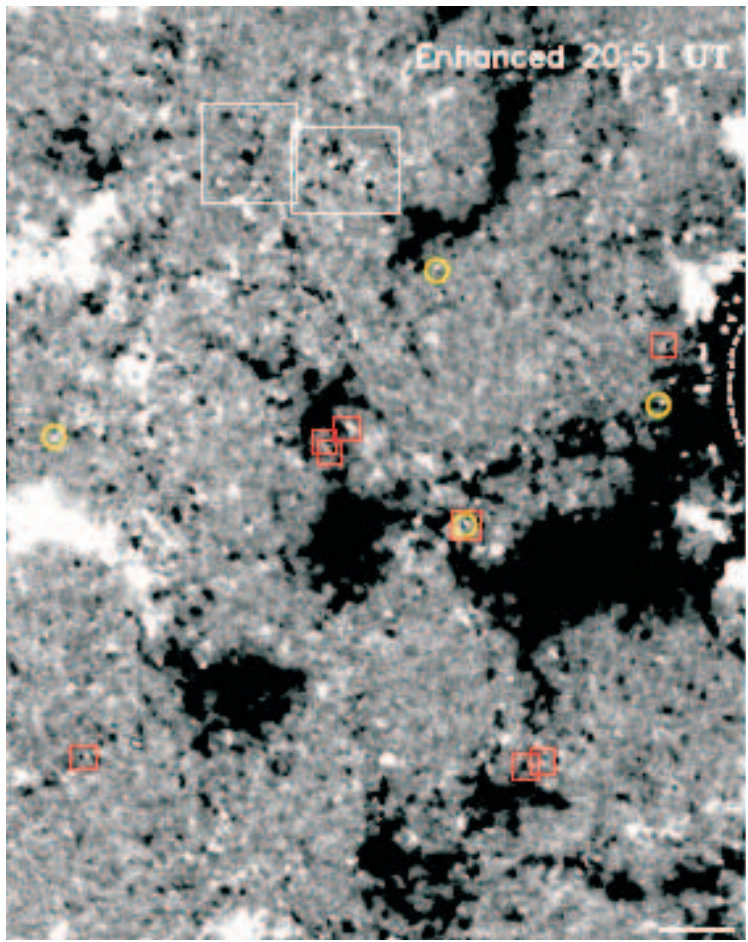}}
\caption{The SOT/NFI line-of-sight magnetogram at 20:50:43 UT, 11
December 2006, is drawn in the same style as for Figure 1. A bar in
the lower-right corner indicates a 10 arcsec length. The box framed
by dashed lines shows an area of cluster flux emergence, while an
adjacent box drawn with dot-dashed lines shows a flux convergence
center. Red boxes and yellow circles have the same meaning as in
Figure 1. }\label{fig2}
\end{figure}

Extremely rich information about IN bipolar appearance, evolution
and disappearance is exhibited by {\it Hinode} SOT/NFI observations,
though it is not easy to describe. In this study we try to explore
and show the most conspicuous behavior of bipoles during their
appearance, not revealed by earlier ground-based observations with
lower spatial resolutions. A few sub-fields in each of the
magnetograms in the chosen time intervals, selected rather randomly,
are used for our exploratory work. We leave for further studies a
more complete statistic description of the properties of the flux
evolution of IN magnetic elements.

\subsection{Calibration of flux measurements}

Following the procedure suggested by Chae et al. (2007), we
calibrate the magnetic flux measurements by comparing the SOT/NFI
observations adopted in this study, with quasi-simultaneous SOT/SP
observations. For each of the four quiet regions, we selected a
subfield for which NFI and SP observations were taken
quasi-simultaneously. For each subfield to match temporarily the two
types of observations, we created a combined NFI magnetogram with
stripes taken from the magnetograms at different times which are as
close as possible to the scanning interval of Stokes V profiles in
SP observations. In this way, the observed magnetic structures are
present almost at the same time in both NFI and SP observations (see
the top panels of Figure 3). In the present paper, we use the same
calibration and procedure as in Zhou et al. (2010).

\begin{figure}
\centerline{\includegraphics[width=0.9\textwidth,clip=]{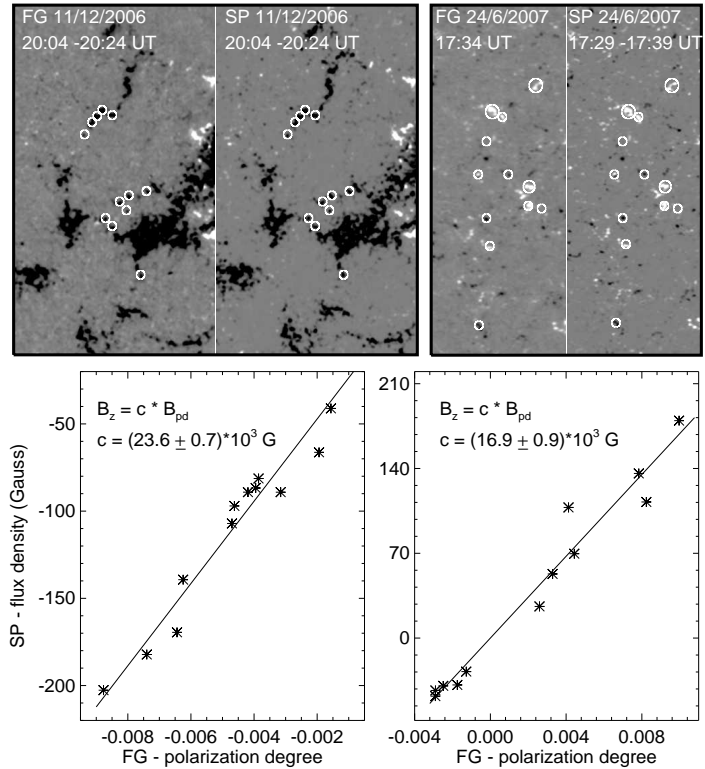}}
\caption{Top panels: quasi-simultaneous SOT/NFI and SP line-of-sight
magnetograms for selected windows. The SP magnetograms are
reconstructed from an inversion of Stokes V profiles using an ME
atmospheric model. The SP magnetogram has been scaled to the dynamic
range of $\pm 300$ G; while for that of SOT/NFI, the dynamic range
of V/I is set to $0.014$. From each pair of magnetograms, a few
well-defined NT elements are selected and encircled. Bottom panels:
correlation of SOT/NFI V/I and the inverted flux density from SP V/I
profiles. The slope found for each set of observations is used to
calibrate the NFI magnetograms.}\label{fig3}
\end{figure}

As demonstrated by Chae et al. (2007), for flux densities $<$ 1 kG a
linear relationship exists between the observed flux density in NFI
observations and the `true' flux density, obtained from the
inversion of SP data. Therefore, it is meaningful to select a few
well-defined NT magnetic elements and obtain a calibration constant
by forcing the measured flux of these elements to be the same in
both types of observations. The relationships for both ENT and QNT
are shown in the bottom panels of Figure 3. For the observations of
ENT regions taken in the FeI 6302 {\AA} line, the calibration
coefficient falls in the range of 23600$\pm$700 G, while for the
observations of the QNT region in the NaI 5896 {\AA} line, the
calibration coefficient is 16900$\pm$900 G. The calibration
coefficient is higher than for other filter-based magnetographs
operated on the ground (see Wang et al., 1996), but consistent in
order of magnitude. The magnetograms taken in NaI 5896 {\AA} appear
to be more sensitive than those taken in FeI 6302 {\AA}.

\subsection{Apparent oscillation of the IN flux density}

There are very clear signals of oscillations in the apparent flux
density observed in the IN regions, with a magnitude of $\pm$1 G
around the mean background flux density. In Figure 4, the average
net flux density of a very quiet IN area within a typical network is
shown.

\begin{figure}
\centerline{\includegraphics[width=0.75\textwidth,clip=]{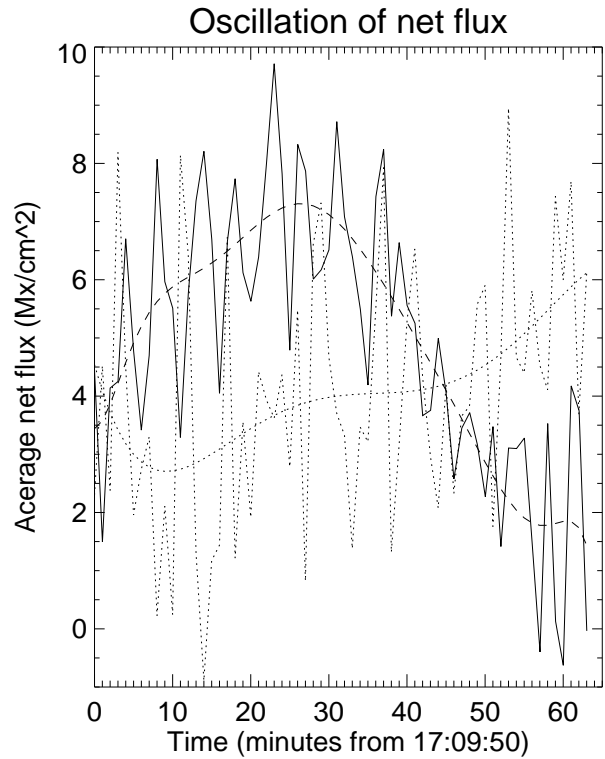}}
\caption{Mean net flux density within a network marked by a circle
in Figure 1. The solid line is the measured mean net flux density,
with the fitted curve indicating the general trend of the background
flux density changes with time. The dotted line represents the
standard deviation of the mean net flux density. To include the
plots in the same figure, 34.0 G were subtracted from the standard
deviation curve.} \label{fig4}
\end{figure}

Within approximately 63 minutes, 15 cycles of flux density variation
appear, indicating an oscillation with a period of 4.2 minutes. At
present, it is not clear whether this oscillation of the net flux
density is a real magnetic oscillation, i.e., of magnetic origin, or
the manifestation of Doppler remnants in the magnetic flux
measurements.

Although the magnitude of the oscillation, as we measured for many
network cells, is no more than $\pm$ 1-2 G, the average net flux
density changes with time; see the fitted curve for the net flux
density in Figure 4. This would mean that there is a changing
background polarization in the flux measurements, which affects
those for weak magnetic features. Moreover, the oscillation would
cause mis-identification of tiny weak magnetic elements. Therefore,
in this set of studies, whenever necessary, we perform a temporal
average to reduce the oscillation effects.

\section{Magnetic bipole appearance}
\subsection{In clusters of mixed polarities}

The analysis of {\it Hinode} data has fully confirmed earlier
observations of a flux emergence center which appears somewhere
within the network (Wang et al., 1995). In such an emergence center,
IN flux appears in the form of a cluster of mixed polarities.

An example of a flux emergence center in the QNT region is shown in
Figure 5 (see also Figure 1). In the upper-left of the figure there
is an emergence center, where a cluster of mixed polarities started
to show up from 17:31 UT. The cluster reached its maximum
development at 17:46 UT and faded after 17:50 UT, suggesting a
lifetime of about 30 minutes. We have clearly observed that in this
cluster of mixed polarities, there were, at least, 4 ephemeral
regions (ERs) which are marked in the figure by green ovals. Their
flux density, size, and location fall into the typical categories of
IN magnetic elements. In this paper, we call them as intranetwork
ephemeral regions (IN ERs). Two of the IN ERs showed the same
magnetic axis orientation, suggesting one or two co-aligned flux
bundles emerging in the convective plasma. The maximum unsigned flux
of each ER in the cluster is (5.9$\pm$2.0)$\times 10^{17}$ Mx in
average, and the separation of the two polarities, 3.3$\pm$2.1
arcsec. The positive flux of these ERs is dominant in this cluster.

Since 17:35 UT, another cluster appeared in the lower-right part of
the field of view (FOV). Three larger ERs, each with total unsigned
flux larger than $10^{18}$ Mx, are identified and marked by green
ovals. They were generally co-aligned, indicating the emergence of a
bundle of magnetic flux from below. Flux cancellation within the
cluster seem to suggest a serpentine structure of the flux bundle
near the surface (see Cheung et al., 2008). However, it is noticed
that in SOT/NFI observations the small-scale emerged flux is always
in the form of point-like flux patches instead of sheets. From
ground-based observations of the highest spatial resolution, the
smallest magnetic field structures appear to arise from isolated
points in the intergranular lanes, instead of arising as continuous
flux sheets confined to the lanes (Goode et al., 2010).

\begin{figure}
\centerline{\includegraphics[width=0.75\textwidth,clip=]{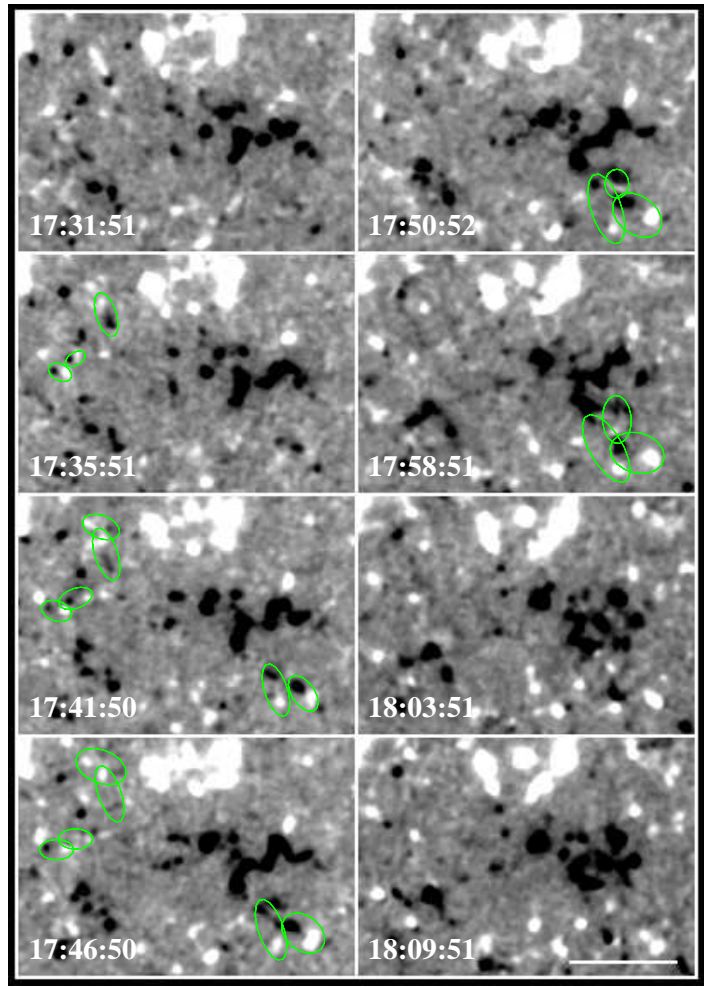}}
\caption{Time-sequence of magnetic flux evolution in the flux
emergence center framed with dashed lines in Figure 1. A bar in the
lower-right corner indicates a length of 10 arcsec. An emergence
center in the upper-left corner appeared since 17:31 UT and faded
after 17:50 UT. In the cluster of mixed-polarity elements, 4 IN ERs
are marked by green ovals. Another cluster emergence started later
at 17:35 UT in the lower-right corner of the FOV, in which 3 larger
ERs were identified.}\label{fig5}
\end{figure}

A flux emergence center within a network in the ENT region is shown
in Figure 6 (see also Figure 2). A cluster of mixed polarities first
appeared at 20:10 UT in the lower-left of the FOV and flux emergence
extended later to the middle and upper parts. In this cluster 4 IN
ERs are clearly identified. Its maximum development took place at
approximately 20:20 UT. In the decay phase of the cluster, one ER
grew up and became a relatively large ER with total unsigned flux
close to $10^{18}$ Mx by 20:32 UT. Moreover, another ER emerged in
the same location and showed the same magnetic orientation, even
though the cluster had already disappeared. The lifetime of this
cluster was about 30 minutes. As for the cluster described in Figure
5, most of the identified IN ERs in this cluster have roughly the
same magnetic orientation. Again, another cluster started to emerge
at 21:04 UT in the upper-right corner of the FOV, in which at least
one or two IN ERs show up clearly.

\begin{figure}
\centerline{\includegraphics[width=.75\textwidth,clip=]{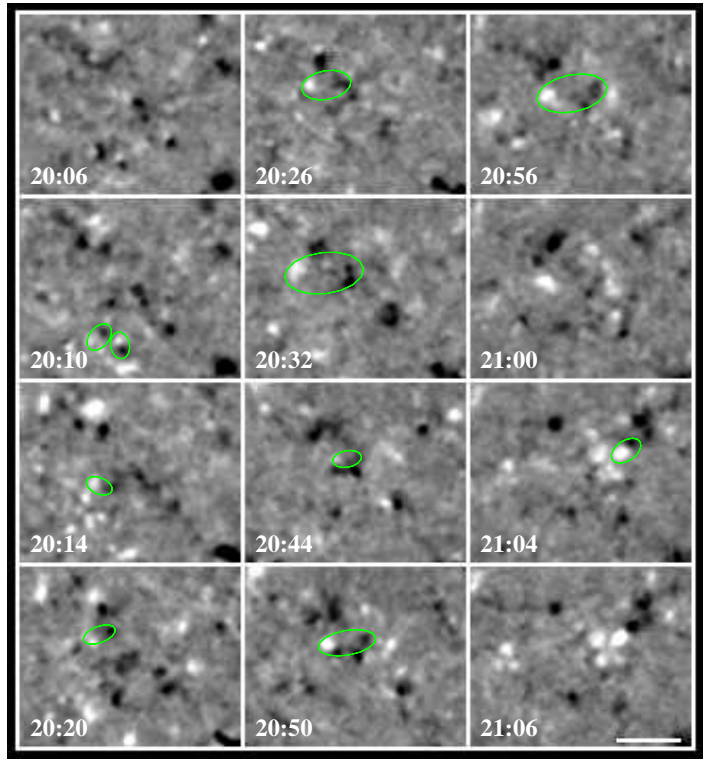}}
\caption{Time-sequence of the magnetic flux evolution in the flux
emergence center framed in Figure 2 with dashed lines. A bar in the
lower-right corner indicates a 10 arcsec length. An emergence center
in the lower left corner appeared at 20:10 UT and faded after 20:26
UT. In this cluster of mixed polarities, 4 IN ERs are marked by
green ovals. Later at 21:04 UT another cluster started to emerge in
the upper-right corner of the FOV. One ER was clearly seen and
marked by a green oval.}\label{fig6}
\end{figure}

In earlier ground-based observations (Wang et al., 1995), we
occasionally found small bipoles, {\it i.e.}, ERs, could be
identified in the cluster of mixed polarities. With {\it Hinode}
SOT/NFI data, the largely improved spatial and adequate temporal
resolutions enable us to reveal that in each of the clusters there
are always a few well-defined ERs. Most of the IN ERs in a cluster
have a total unsigned flux less than $10^{18}$ Mx, while the total
unsigned flux of each cluster is several times $10^{18}$ Mx. The
cluster behavior, {\it e.g.}, the rather ordered alignment of mixed
polarities and a few clearly identified ERs, seem to be consistent
with the predicted picture of the serpentine field lines that emerge
into the photosphere in magneto-convection simulations (see Cheung
et al., 2008 and the references therein). However, the total flux
for each well-observed cluster is much less than the one in the
simulation, {\it e.g.}, from several times $10^{19}$ Mx to $10^{20}$
Mx. The observed clusters seem to exhibit interlaced and braided
features as in an emerging bundle of magnetic flux. However, even
with one minute temporal resolution and 0.3 arcsec spatial
resolution, we are unable to identify the magnetic connectivity of
all the mixed polarity elements in an observed cluster.

It is interesting to see that sometimes within the same network, a
flux convergence center is just located in adjacent to a flux
emergence center (see the one framed by a dash-dotted line in Figure
2). The flux convergence is shown in Figure 7. It is not difficult
to see the general convergence of magnetic elements toward to the
center of the FOV. For the two negative elements which converged
toward the center of the FOV and merged into an element, the average
convergence speed was about 4.0 km$s^{-1}$ from 20:22 to 20:52 UT.
Flux convergence is usually achieved by converging motions and flux
cancellation of IN elements of opposite polarity.

In cases of flux coalescence of the same polarity flux patches, we
always observe their cancellation with opposite polarities between
them. The isolated merging of same polarity patches without flux
cancellation with surrounding opposite polarities has never been
found. This fact was previously recognized in our early ground-based
observations ({\it e.g.}, Wang et al., 1995).

\begin{figure}
\centerline{\includegraphics[width=0.6\textwidth,clip=]{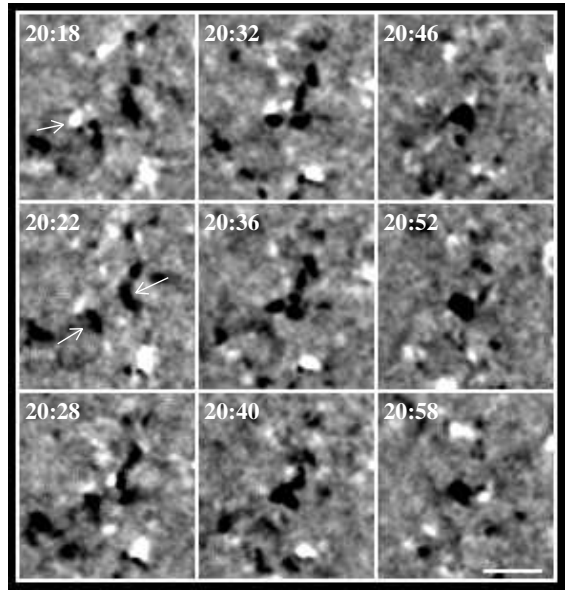}}
\caption{Time-sequence of magnetic evolution in the flux convergence
center framed with a dot-dashed line in Figure 2. A bar at the
lower-right corner indicates a 10 arcsec length. Arrows in the
figure mark 3 elements which moved rapidly to the center of the FOV
with a speed above 4 km$s^{-1}$.}\label{fig7}
\end{figure}

\subsection{Intranetwork ephemeral regions}

For the sampled QNT and ENT regions, lots of tiny ERs within the
cells of the magnetic network are identified. They typically have
the size and flux of IN magnetic elements. Hereby we refer to them
as intranetwork ephemeral regions (IN ERs). A statistical analysis
of IN ERs is undertaken.

\begin{figure}
\centerline{\includegraphics[width=.85\textwidth,clip=]{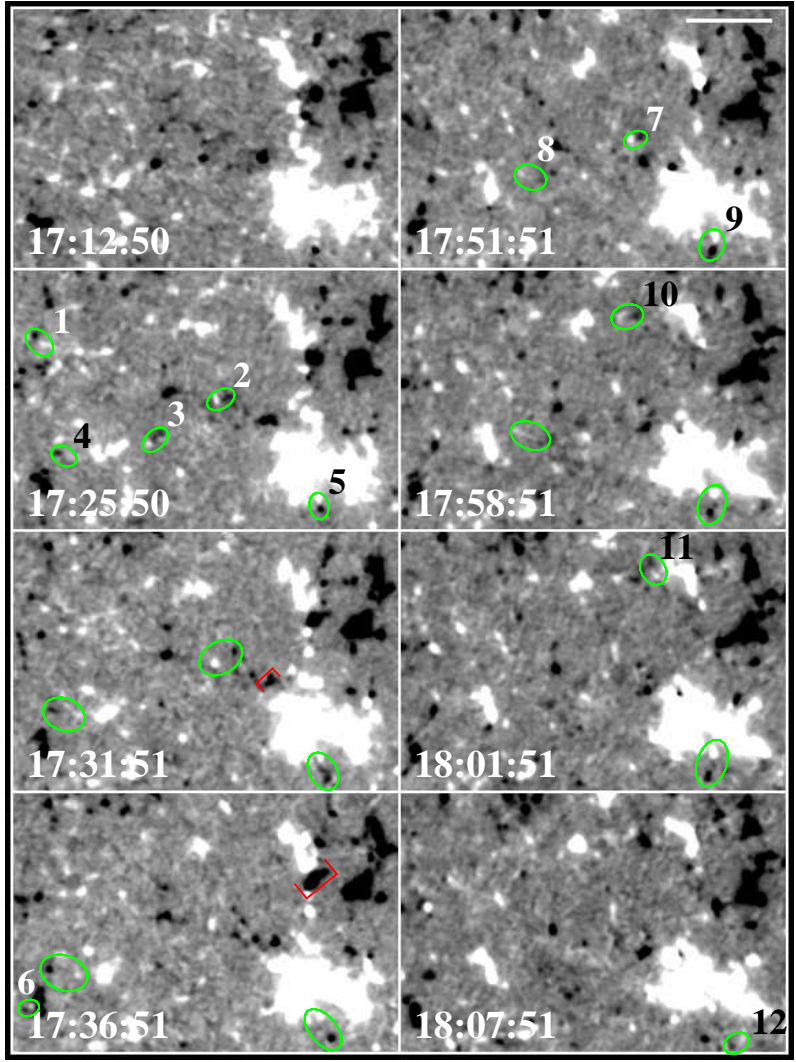}}
\caption{Illustrations of IN ERs for the area of a network cell
framed in Figure 1 by the solid lines. A bar in the upper-right
corner indicates a 10 arcsec length. Twelve IN ERs emerged in this
area, each of which is marked by a green oval in the figure, from
17:13 to 18:08 UT.}\label{fig8}
\end{figure}

In an area of 45$\times$31 arcsec$^{2}$, approximately the size of a
supergranular cell, and in an interval of 55 minutes (Figure 1) we
identified 12 likely ERs. They are numbered in Figure 8 according to
the time of first appearance. Most of them are not difficult to
follow in the successive frames in the figure. Their magnetic axis
distribution is random. The average maximum unsigned flux,
separation, and lifetime of the 12 IN ERs are (7.8$\pm$4.1)$\times
10^{17}$ Mx, 3.3$\pm$1.4 arcsec, and 14$\pm$8 minutes, respectively.
The ERs emerged close to strong NT elements tend to have more
magnetic flux. In Figure 8, ERs 5 and 9 emerged in the strong
network boundary and close to strong positive NT elements. Their
total unsigned fluxes were 1.48 $\times 10^{18}$ Mx and 1.41$\times
10^{18}$ Mx, respectively, which are still two orders of magnitude
smaller than those first reported by Harvey and Martin (1973), and
one order of magnitude smaller than the average flux found by
Heganaar (2001). Careful statistical work needs to be done to
distinguish (if possible) between IN ERs and other `normal' ERs as
described in the earlier literature.

In Figure 8 we mark two cancelling magnetic features with red square
brackets at 17:31 and 17:36 UT, respectively. Two adjacent flux
patches of opposite polarity are not necessarily ERs, but often
pseudo ERs though they are paired in the close vicinity (Martin et
al., 1985). For example, at 17:36 UT, a negative NT element (marked
with a red square bracket) and the adjacent positive NT element of
about the same size looked like an ER. However, looking at their
evolution, we find that they moved toward one another, the magnetic
gradient between the opposite polarities increased, and their flux
disappeared mutually. This is a typical cancelling magnetic feature
(CMF) consisting of two NT elements of opposite polarity (Livi,
Wang, and Martin, 1985; Wang et al., 1988). An IN element of
negative polarity, which was marked at 17:31 UT in the figure,
intruded into the strong positive network elements and cancelled
with them. By 18:01 UT it was encircled entirely by NT positive
flux, and disappeared completely after 18:07 UT. The scenario is
very typical of flux cancellation between IN and NT elements.

\begin{figure}    %%%%%%%%%%%%%%%%%% FIGURE 2
                                % includes the two top panels
   \centerline{\hspace*{0.015\textwidth}
               \includegraphics[width=0.42\textwidth,clip=]{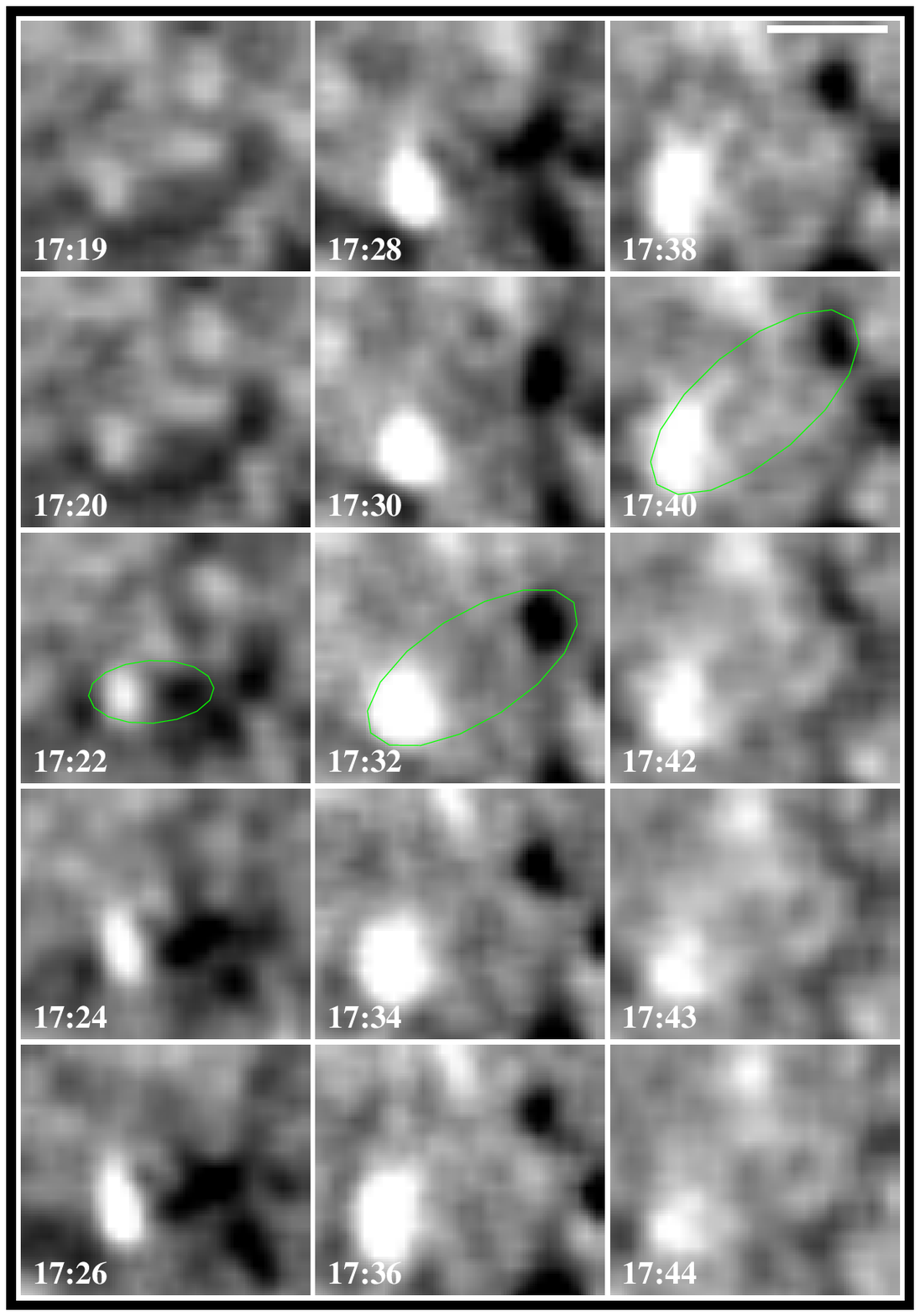}
               \hspace*{-0.02\textwidth}
               \includegraphics[width=0.55\textwidth,clip=]{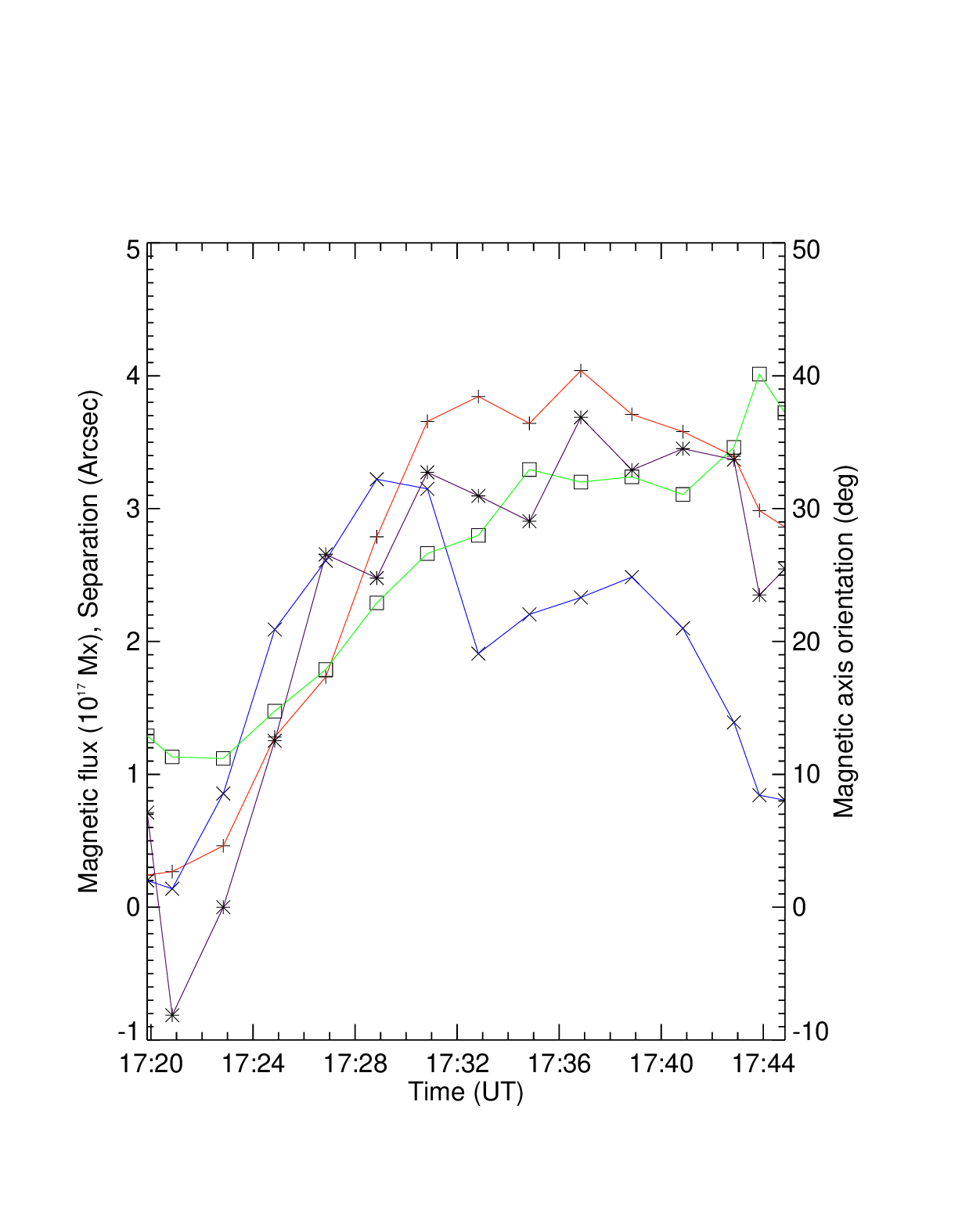}
              }
     \vspace{0.01\textwidth}   % Shift close to the panel top
     \centerline{\Large \bf     % Includes the labels (here needs the color
                                %   package, see beginning of this file)
      \hspace{0.21 \textwidth}  \color{black}{(a)}
      \hspace{0.41\textwidth}  \color{black}{(b)}
         \hfill}
     \vspace{0.005\textwidth}    % Shift back to the panel bottom
\caption{Panels (a): illustration of ER2 from the earliest
appearance to the disappearance of its negative polarity. The bar in
the upper-right corner indicates a length of 2 arcsec. Panel (b):
curves indicating the evolution of the magnetic flux, separation,
and magnetic orientation of ER2. The positive flux is drawn in red
and marked by Plus symbol, negative flux, in blue and by Cross
symbol. The separation of the two polarities is shown in green and
by Square symbol, and the magnetic orientation from the positive to
the negative polarity in dark purple and by Asterisk symbol. Note,
the flux, separation, and orientation have different scales though
they share the same ordinate. The magnetic flux is expressed in
units of $10^{17}$ Mx, the separation in units of 1 arcsec, and the
orientation in units of 10 degrees being the angle positive for an
anti-clockwise rotation starting from the solar equator.}
   \label{fig9}
   \end{figure}

The appearance of ER2 is shown in Panels (a) of Figure 9, and the
flux evolution, separation, and magnetic orientation of its
polarities are shown in Panel (b). Although being a simple ER, it
had a complicated interaction with the surrounding magnetic flux.
Both poles of ER2 did not fully grow while cancelling with nearby
opposite polarities. The negative pole seemed to be clustered with
nearby flux of the same polarity and embodied in the unresolved flux
of opposite polarity with very low flux density. It never grew as
strong as its positive counterpart. The most rapid separation,
growth and rotation took place from 17:24 to 17:30 UT in the early
phase of flux emergence. It is remarkable that this IN ER rotated
anti-clockwise for more than 30$^0$ with a rotating rate of
approximately 0.08$^0$ s${-1}$ (see Panel (b) of Figure 9). The fast
magnetic axis rotation is co-temporal with the flux growth. The
evolution of these characteristics of ER2 are common to other IN ERs
in this study.

Moreover, even with a 1 minute cadence, it is not straightforward to
identify the same ER in two successive {\it Hinode}/NFI frames. The
identification of ERs is affected  by problems due to a low temporal
resolution in three cases: 1) an ER with a shorter lifetime, 2) an
ER subject to  complicated interactions with the surrounding
magnetic elements, and 3) an ER displaying convective motions.
Often, the identification is difficult and uncertain, as in the case
of ER5 and ER9. We are not sure if they are a single ER or two ERs,
or even some peculiar flux appearance. We will discuss them in the
next section.

According to the traditional definition for an ER, the two flux
patches with opposite polarity emerge simultaneously or one
following the other, and grow and separate co-temporally (see Harvey
and Martin 1973). {\it Hinode} data extend the identification of ERs
to those with much lower flux. It has long been believed that ERs
represent small-scale bipole loop emergence (see Parker 1984). The
ERs shown here are the same as the emerging small-scale magnetic
loops, described observationally by Centeno et al. (2007) and
Martinez Gonzalez and Bellot Rubio (2009). In Figures 1 and 2, we
marked a few tiny IN ERs by green circles. They all meet the above
definition for an ER and their emergence can be traced at least in
three successive frames. The ERs on the quiet Sun may have a total
unsigned flux less than 10$^{17}$ Mx and separation less than 1
arcsec.

\begin{figure}
\centerline{\includegraphics[width=0.9\textwidth,clip=]{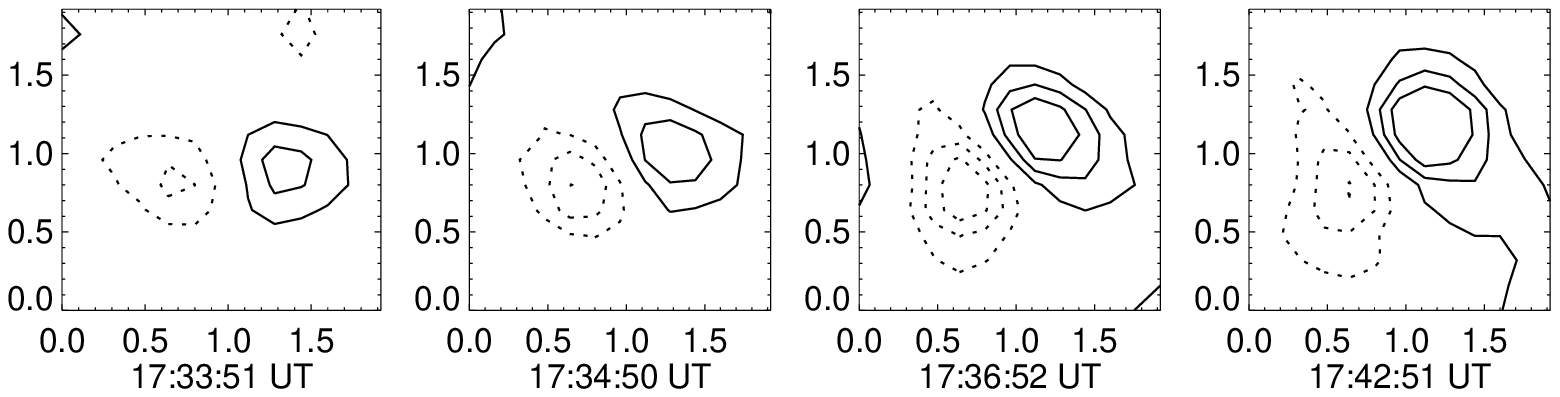}}
\centerline{\includegraphics[width=0.95\textwidth,clip=]{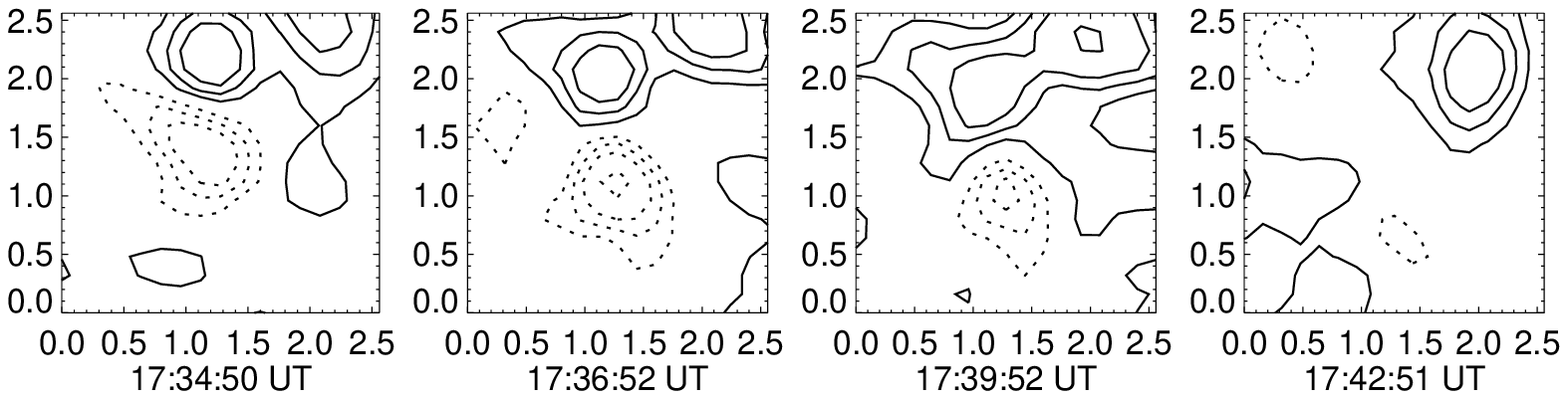}}
\centerline{\includegraphics[width=0.95\textwidth,clip=]{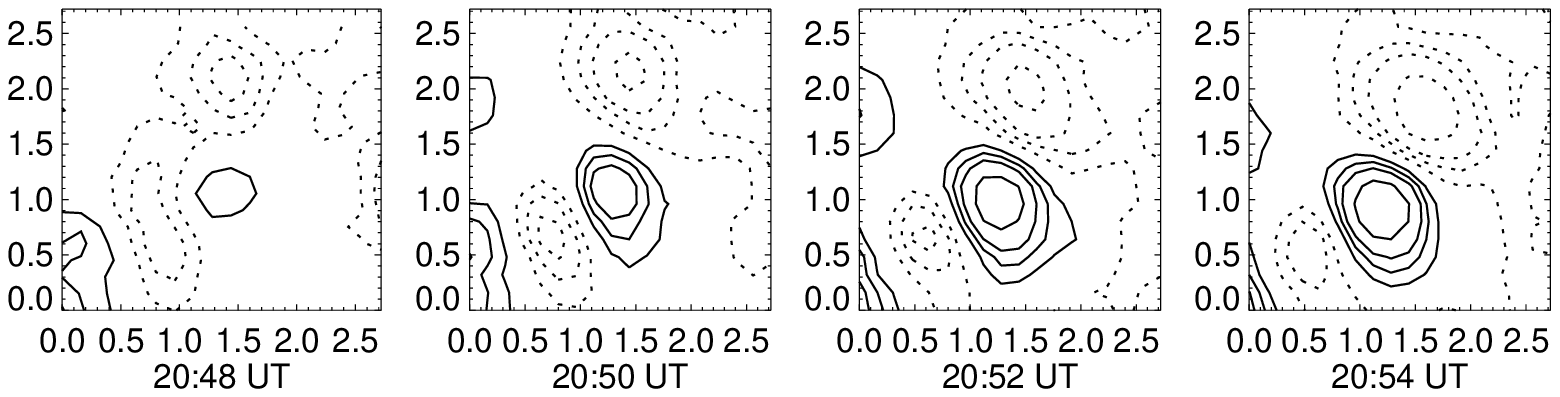}}
\caption{Top panels: a tiny ER in shown by contours of the flux
density. Its total unsigned flux is about $1.0\times 10^{17}$ Mx and
the separation less than 1 arcsec. Middle panels: a tiny CMF is
shown by contours of the flux density. Bottom panels: a tiny ER
shown by contours of the flux density. The axes are in arcsec. The
contour levels are $\pm$10, 20, 30, 50 G, respectively.
\label{fig10}}
\end{figure}

In the top panels of Figure 10, the emergence of a tiny ER (marked
as the uppermost green circle in Figure 1) is shown by contours of
the magnetic flux density. It fitted exactly the definition of an ER
and showed imbalance magnetic flux between opposite polarities in
its later development. The flux imbalance of this ER resulted from
the merging of its positive pole with a nearby positive flux.
Moreover, its magnetic axis rotated anti-clockwise for more than 20
degrees in 5 minutes. In the bottom panels of Figure 10, another
tiny ER (marked by a green circle in the central part of Figure 2)
is shown in the same way. Its positive pole is in the center of the
FOV and the negative one to its north. As usual, this ER exhibited
flux imbalance between opposite polarities. The imbalance is caused
by the cancellation of the ER's positive pole with the negative flux
to its south. The positive pole grows and simultaneously cancels
with a nearby negative flux patch. As the result of flux
cancellation, the negative flux patch in the south showed a rapid
decrease both in area and flux density and the ER positive pole
never grew as strong as the negative one. It is important to notice
that even at the smallest spatial scale, IN ERs follow the typical
scenario in emergence as larger ERs (Schrijver and Zwaan, 2001;
Thomas and Weiss, 2008). Often, an apparent bipole with
approximately the same flux and size, e.g, the apparent bipole at
20:50 UT in the lower-left of the FOV in the bottom panel, is not an
ER. The automatic selection of ERs from magnetograms with poor
temporal resolution should be considered with great cautions (see
Hagenaar, 2001).

A few tiny CMFs are marked in Figures 1 and 2 by small red squares.
One of them is shown in the middle panels of Figure 10 (marked by a
small red square in the middle-right of Figure 1). Its negative pole
is in the middle of the FOV. The positive pole grew from 17:35 to
17:40 UT by coalescence with a positive polarity to its north-west.
The cancellation, as usual, started first at the place where there
were high field gradients. As the result of flux cancellation, the
negative flux fragmented into two pieces, one of them was very
small. The positive flux tended to move toward the cancellation
site. From the curvature of the periphery of the positive flux, we
know where flux loss had taken place. The negative flux disappeared
completely from 17:40 to 17:43 UT. The scenario is very much typical
of any CMF though happening between tiny flux patches of opposite
polarity.

\subsection{Peculiar flux appearance and magnetic float}

Observations of flux submergence have been reported for ARs using
early ground-based observations (Wallenhorst and Topka, 1982; Rabin
et al., 1984; Zirin, 1985; Harvey et al., 1999; Kalman, 2001). These
observations sometimes were considered to be questionable as the
spatial and temporal resolutions were not high enough and in most of
the cases no vector field observations were available in the
analysis. Yang et al.(2009) reported an example of flux submergence
of an ER in a coronal hole observed with {\it Hinode} high spatial
resolution. It is worthwhile emphasizing that the term of flux
``submergence'' implies the physical scenario of an $\Omega$ loop
that once broke out into the photosphere from below and now
retracted into the sub-photosphere again. Without 3-D magnetic
observations, including data from the sub-photospheric layers, it is
really hard to believe that the observed submergence is real.
Therefore, the term ``submergence'' is used in this paper to
describe that the opposite polarities of an ER reverse their growth
and separation.

\begin{figure}
\centerline{\includegraphics[width=0.95\textwidth,clip=]{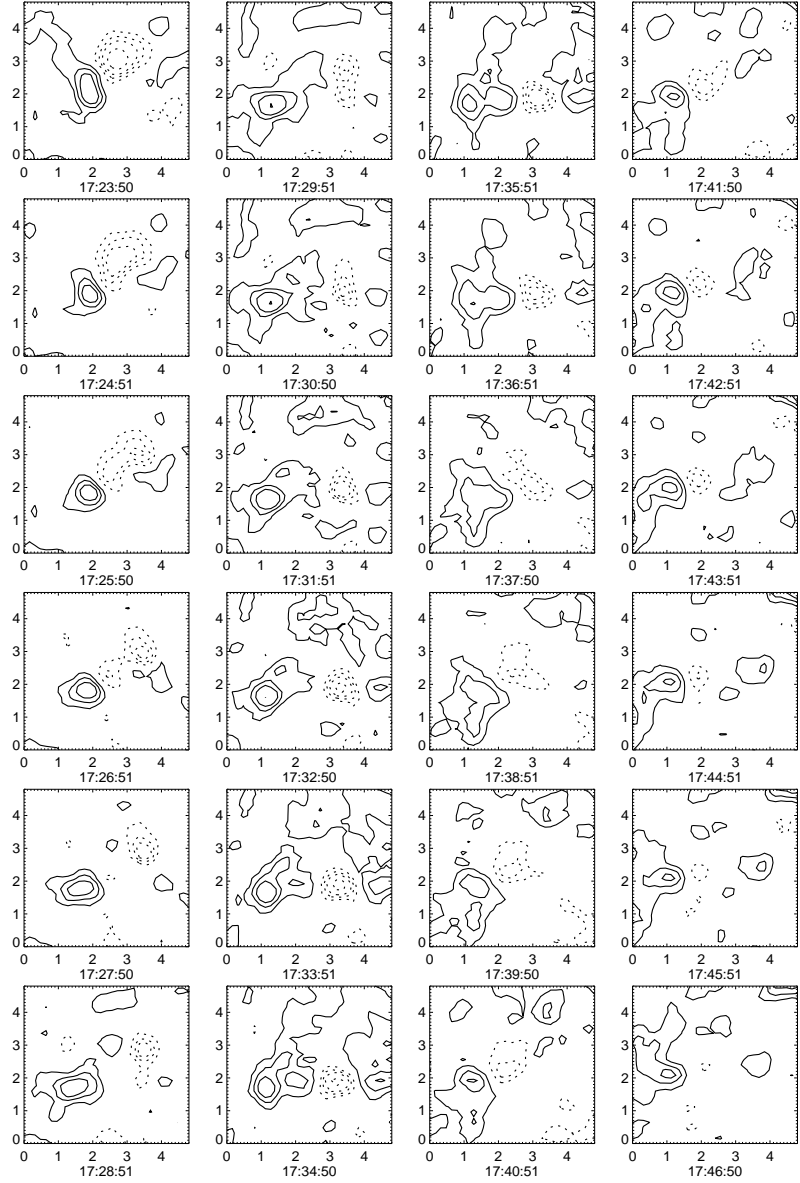}}
\caption{Time evolution of an ER, ER 3 in Figure 8, shown by contour
maps of magnetic flux density. The axes are in arcsec, and the
contour levels are $\pm$10, 20, 30, 50 G, respectively.}
\label{fig11}
\end{figure}

Among all the ERs described above, only ER3 in Figure 8 exhibited
the unusual scenario of flux emergence followed by an apparent
submergence. In Figure 11, we show the whole evolution of ER3 with
contour maps at one minute cadence. In this way, the quantitative
evolution can be viewed more clearly. The ER seemed to emerge
partially in pre-existing flux. From the earlier appearance of the
ER at 17:23 UT to the disappearance of its negative polarity at
17:46 UT, the evolution can be divided in steps: (1) clearing up the
bipole emergence from 17:24 to 17:28 UT, during which a portion of
the negative pole separated and cancelled with the pre-existing
positive pole while moving outward; (2) evolving to its maximum
development at 17:34 UT when its negative pole became strongest in
flux density and its positive pole showed up the secondary flux
center (with a contour value of 20 G); (3) setting up for
submergence in the interval of 17:34 to 17:41 UT, during which its
positive pole had its secondary flux patch cancelled with the ER
negative pole and its positive flux coalesced; (4) shrinkage of the
bipole from 17:41 to 17:47 UT. The first and third steps are
complicated and involve interior cancellation between pieces of flux
of two opposite polarities and external cancellation with
surrounding opposite polarities. After these two steps, the ER
seemed to strip the surrounding flux and appeared as a more simple
bipole. It is interesting to notice that the magnetic axis of the
overall bipole rotated clockwise in the first two steps for about 30
degrees. Then in step 3, it rotated back anti-clockwise, and finally
in step 4, the axis rotated clockwise again close to east-west
direction. From this and other examples, we have learned that the
magnetic axis of an ER always rotates rapidly during rapid flux
growth and shrinkage.

To our knowledge, this is the first evidence of a bipole emergence
followed by apparent submergence, {\it i.e.}, an ER cancelling
itself. We believe that there is a bipole emergence and submergence
because we track the two opposite polarities continuously at high
enough cadence and spatial resolution.

Although it is not often to see an ER cancelling itself, quite a few
examples show the opposite process, a cancelling magnetic feature
that finally reverses its evolution and appears as an ER. In Figure
12, we show a pair of opposite polarities approaching and cancelling
from 17:42 - 17:53 UT. However, after the opposite polarities
appeared to come into contact, they reverse their evolution. A pair
of polarities with almost the same flux started to grow and separate
like a usual ER. This can be seen until the negative flux
disappeared. For comparison, a normal ER is surrounded with a green
oval in the first three magnetograms. The ER grew and separated
during the whole time-sequence, a very clear example of ``direct
emergence'' (see Toriumi and Yokoyama, 2010).

\begin{figure}
\centerline{\includegraphics[width=0.6\textwidth,clip=]{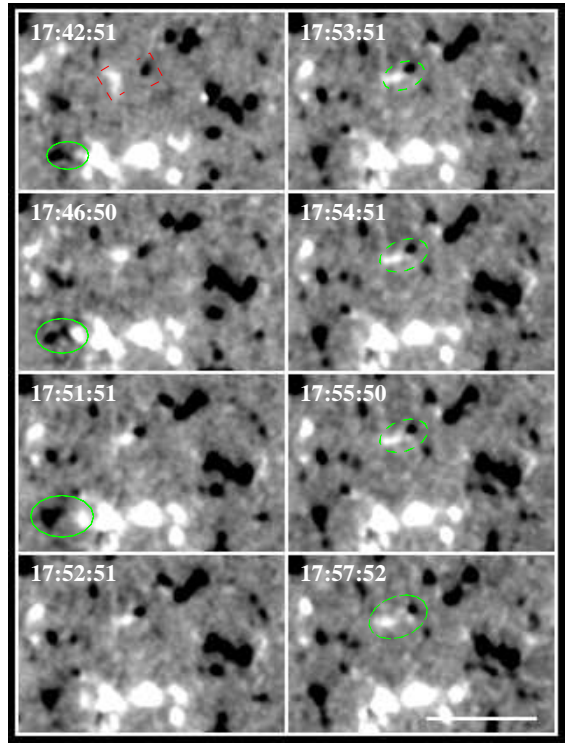}}
\caption{A cancelling magnetic feature is marked by open square
brackets at 17:42 UT. After approach and cancellation of opposite
polarities for 10 minutes, the bipole, marked by a dashed green
oval, appears to separate from 17:53 UT as a usual ER. To compare,
an ER at the NT boundary is shown by a solid green oval. A bar at
the lowest right corner indicates a length of 10 arcesec.}
\label{fig12}
\end{figure}

Sometimes, a bipole is seen to shrink first, then, to grow for some
time, and to shrink again. In the process of
shrinkage-growth-shrinkage its magnetic axis rotates greatly. An
example of this type of flux evolution is shown in Figure 13. The
first shrinkage took place from 17:30 to 17:42 UT. The shrinkage was
followed by growth from 17:42 to 17:50 UT. The bipole shrank again
after 17:50 UT. The negative polarity of the bipole disappeared
after 17:54 UT. The flux decrease and increase in the negative pole
was companied by the co-temporal decrease and increase in the
positive polarity. During this evolution, the magnetic axis rotated
clockwise from about 40 degrees to -80 degrees. The negative pole
rotated around the stronger positive pole for more than 100 degrees
with a rotating rate of 0.1 degrees/s (for rapid sunspot rotation
see Zhang, Li, and Song, 2007 and Yan et al., 2009). The bipole
shown here looks like a \emph{magnetic float} in the convective
photosphere, {\it i.e.}, the $\Omega$-loop is observed to sink,
rise, and sink again while rotating.

\begin{figure}
\centerline{\includegraphics[width=0.7\textwidth,clip=]{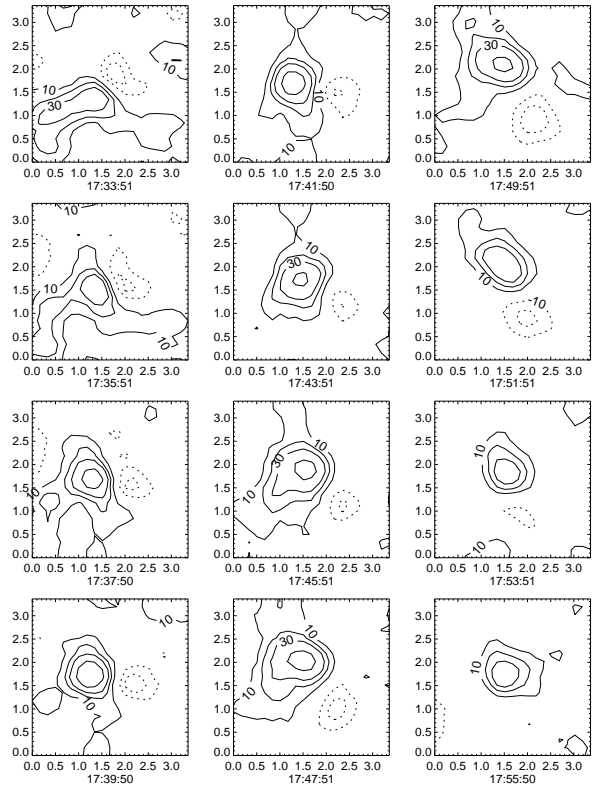}}
\caption{A bipole showing shrinkage-growth-shrinkage during a time
interval of 20 minutes. } \label{fig13}
\end{figure}

The evolution of ER5 and ER9 in Figure 8 is complex (see Section
3.2). Observationally, analyzing a short time interval with an
intermittent image sampling, we would have no doubts that their
evolutions were consistent with two ERs. However, if we viewed the
evolution in a longer interval and with higher cadence, we would
find that the evolution was very complicated (see Figure 14). We
could not be sure if we did observe two ERs or just one ER with a
peculiar evolution.

First, we saw two opposite polarity elements which moved closer from
17:20 to 17:25 UT while their flux slightly decrease, {\it e.g.},
characteristic of a ``cancelling magnetic feature''. Secondly, from
17:25 to 17:34 UT, we basically saw the separation and some flux
increase of these two opposite polarities, although interacting with
other smaller magnetic elements. Even with one minute cadence, we
could not ascertain if the negative polarity that we saw at 17:34 UT
was the same as that at 17:25 UT. Thirdly, we could only tentatively
conclude that only an ER (marked as ER5 in Figure 8) was observed.
Notice that from 17:27 to 17:31 UT additional negative flux patches,
whose magnetic connectivity was hard to tell, appeared. Fourthly,
from 17:34 UT, we seem to see another cancelling magnetic feature,
{\it i.e.}, the approach and shrinkage of opposite polarities until
17:48 UT. Finally, another ER (marked as ER9 in Figure 8) was seen
from 17:48 to 18:09 UT. The fast separation of opposite polarities
seemed to really represent a buoyant flux emergence with an increase
of the magnetic field strength.

\begin{figure}
\centerline{\includegraphics[width=0.65\textwidth,clip=]{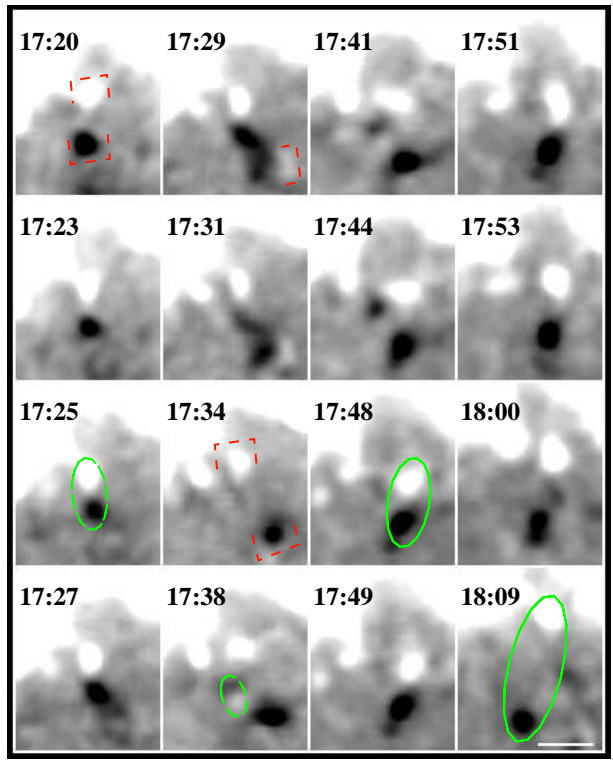}}
\caption{A bipole showing double shrinkage-growth during an hour.
The bar in the lowest-right corner indicates a length of 2 arcsec.}
\label{fig14}
\end{figure}

It is interesting to notice that the total flux of this bipole
changed in the range of (8 - 15)$\times 10^{17}$ Mx, {\it i.e.},
within a factor of two. We have not observed a rapid flux emergence
like the one in other ERs or fast CMF-like flux disappearance. We
seem to witness a floating bipole suspended in the dynamic
photosphere, moving up and down, while changing its orientation
slightly. The other examples shown in Figures 11 to 13 also seem to
show a magnetic float, at least, during some interval in their
lifetime.

The magnetic float behavior is quite commonly seen in SOT/NFI
magnetograms with high spacial resolution and adequate cadence. From
another view point, we may be observing the complexity of flux
emergence, ``two-step emergence'' or multiple-step emergence, or
even ``failed emergence'', as suggested in the numerical simulation
by Toriumi and Yokoyama (2010). The magnetic fields in the
complicated flux emergence is not so strong but within the typical
range of intranetwork fields (see Jin, Wang, and Xie, 2011).

The behavior of magnetic floats implies that these IN bipoles are
close to an equilibrium between magnetic and convective drag forces.
It is assumed that the importance of the drag force increases with
decreasing tube radius and that tubes can be dragged even if they
have strong magnetic fields of the order of thousands of Gauss (see
S\'{a}nchez Almeida, 2001 and references herein). However, as Jin,
Wang, and Xie (2011) demonstrated, the IN fields are dominantly weak
in terms of the equipartition field with the convection in the
photosphere. The destiny of a floating bipole is either to emerge
due to a magnetic buoyancy instability when the magnetic field is
enhanced in the flux loop, or to submerge forced by local convection
together with magnetic tension, or breaking up into fragments by
interacting with other magnetic elements.

\section{Conclusions and Discussion}

With the unprecedented spatial resolution of SOT/NFI magnetograms,
we have confirmed the existence of flux emergence centers which
appear somewhere within the network as earlier found (Wang et al.,
1995). IN flux appears in such an emergence center in the form of a
cluster of mixed polarities. It has been further clarified that in
each of the clusters there are always a few well-developed ephemeral
regions, whose magnetic orientation appears to be in general the
same.

The sampled IN ephemeral regions have fluxes from less than 1$\times
10^{17}$ Mx to 15 $\times 10^{17}$ Mx, separations of the two
polarities from 3 arcsec to 4 arcsec, and lifetimes of 15-20
minutes. The smallest IN ERs have a maximum separation between
opposite polarities of less than 1 arcsec, and a total unsigned flux
of a few times of $10^{16}$ Mx.

We have shown that many IN bipoles behave like a magnetic float in
the convective photosphere. They are often found to first shrink and
then grow, or vice versa during a time interval of several tens of
minutes. The scenario of shrinkage-growth and/or growth-shrinkage
implies that these IN bipoles are in a quasi-equilibrium in the
photosphere. They are likely to have an equipartition field strength
with plasma kinematics, and are buffeted again and again by plasma
convection.

The force keeping a magnetic bipole in an equilibrium can be
simplified as
\begin{equation}
F=\frac{B_i^2}{2\mu\Lambda_e}+(\frac{T_i}{T_e}-1)\frac{P_i}{\Lambda_i}
-\frac{B_i^2}{\mu R}+f_c,
\end{equation}
where, as usual, $B$, $T$, and $P$ are the magnetic induction,
temperature, and pressure; $\Lambda$ is the pressure scale height
$\frac{kT}{mg}$ with the Boltzmann constant $k$ (Priest, 1982), and
R is the local curvature radius, or roughly speaking half the
footpoint separation of the bipole loop. The subscripts `i' and `e'
refer to the internal and external loop quantities. The last term in
the equation, $f_c$, is the force of the convective plasma on the
bipole, which is difficult to be described accurately. In the
equation, the rationalised mks units are used, {\it e.g.}, $B$ is in
tesla (T) and $\mu=4\pi \times 10^{-7}N\,A^{-2}$. The behavior of
even a simple bipole in the convective photosphere is too complex
when all the physical forces are considered. Radiative MHD
simulations are needed to clarify the physical picture completely
(see V\"{o}gler et al., 2005).

Without the convective force and the difference between the
temperatures inside and outside the flux loop, the bipole can
submerge if $R<2\Lambda_e$ (see van Ballegooijen and Martens, 1989).
In the photosphere $\Lambda_e$ is about 150 km. For the tiny bipoles
with separations of less than 1 arcsec simple submergence could
account for the bipole disappearance from the photosphere. However,
for a bipole with larger separation of its two polarities, the
convection buffeting would make it temporarily grow and shrink
easily, provided the magnetic field inside is not very strong to
initiate the buoyancy instability. Whenever, the temperature inside
the loop is higher than that in the surrounding, the additional
buoyancy force, {\it i.e.}, the second term on the right hand side
of equation (1), will favor bipole emergence. The internal dynamics
of a small-scale loop and its interaction with the external plasma
makes the bipolar flux appearance in the intranetwork regions to be
complicated.

Cluster emergence of mixed polarities is common and often has a
preferred magnetic orientation, but it is never so ideal as Thornton
and Parnell (2011) have searched for. Current spatial resolution and
cadence in {\it Hinode} observations do not allow a definitive
diagnosis of magnetic connectivity and pairing of opposite
polarities in a cluster. In each of the observed clusters, there
were always a few IN ERs with total unsigned magnetic flux of a few
times of 10$^{18}$ Mx. The morphology of the clusters with mixed
polarities is consistent with magneto-convection simulations (see
Cheung et al., 2008; Toriumi and Yokoyama, 2010). However, the total
magnetic flux in a cluster is much less than that suggested by these
simulations. This seems to hint that the horizontal flux structures
responsible for flux emergence in IN regions are located much
shallower than considered in these simulations, certainly not at the
bottom of convection zone.

The IN ERs found in this study are as small as having a maximum flux
less than 10$^{17}$ Mx, and a separation of the polarities smaller
than 1 arcsec. However, they have all the properties described by
the pioneer studies of Harvey and Martin (1973) and Martin and
Harvey (1979). Their behaviors are also found to be consistent with
space observations of SOHO/MDI (Hagenaar, 2001) and the reports of
other authors based on {\it Hinode}/SOT observations (see Thornton
and Parnell, 2011 and references therein). It is remarkable that
Thornton and Parnell (2011) found a unique power-law distribution of
the flux emergence rate, which spans nearly seven orders of
magnitude from 10$^{16}$ to 10$^{23}$ Mx.

\begin{table}
\caption{The buoyancy effect in bipole emergence. } \label{tbl-0}
\begin{tabular}{rrlrrl}
  \hline
Case & Depth (km) & $\rho_e (kgm^{-3})$ & $B_i (Tesla)$ & T(K) &  $\delta$ \\
  \hline
1 & 20,000 & 0.25 & 0.1  & 2.5$\times 10^{5}$ & 7.7$\times 10^{-6}$ \\
2 & 20,000 & 0.25 & 1.0  & 2.5$\times 10^{5}$ & 7.7$\times 10^{-4}$ \\
3 & 20,000 & 0.25 & 10.0 & 2.5$\times 10^{5}$ & 7.7$\times 10^{-2}$ \\
4 & 20,000 & 0.25 & 0.01 & 2.5$\times 10^{5}$ & 7.7$\times 10^{-8}$ \\
5 & 1,000 & 8.0$\times 10^{-4}$ & 0.1 & 1.5$\times 10^{4} $ &
4.0$\times 10^{-2}$ \\
6 & 1,000 & 8.0$\times 10^{-4}$ & 0.01 & 1.5$\times 10^{4} $ &
4.0$\times 10^{-4}$ \\
7 & 500 & 2.5$\times 10^{-4}$ & 0.01 & 1.0$\times 10^{4} $ &
1.9$\times 10^{-3}$ \\
\hline
\end{tabular}
\end{table}

The fact that buoyant flux emergence can happen in very weak IN
bipoles seems to imply a shallow depth from where the bipoles start
emerging. Following Priest (1982), for a simple case ignoring the
second and fourth terms in Equation (1), we can use $\delta$ to
quantify how important is magnetic buoyancy
\begin{equation}
\delta=\frac{\rho_e - \rho_i}{\rho_e}=\frac{m}{2\mu
k}\frac{B_i^2}{\rho_eT} \simeq48.19\frac{B_i^2}{\rho_eT},
\end{equation}
where the magnetic field $B_i$ is in T, $\rho_e$ in kg m$^{-3}$, and
T in K.. In Table 2, we list the values of $\delta$ for different
magnetic field intensities and initial depths of the emerging
bipole. The cases 2-4 in the table represent the typical conditions
of ``two-step'', ``direct'', and ``failed'' flux emergence in the
simulation of Toriumi and Yokoyama (2010). For the failed emergence
the buoyancy effect is so small that can be ignored. If we admit the
emergence depth to be shallow enough, cases 5-7 could represent the
observed ``two-step'' or ``direct'' flux emergence, although the
magnetic field inside the loop is weak.

In the observations with 1-2 minute temporal resolution and 0.3
arcsec spatial resolution, the magnetic float behavior is found to
be quite common for intranetwork bipoles. This implies that
distinguishing ephemeral regions from cancelling magnetic features
is not an easy task at all. Sometimes, an IN ER suffers various
processes during its emergence. Two-step or multi-step flux
emergence would be easily seen. Failed flux emergence is possible to
be observed too. Their final break out as an ER requires the
increase of field inside the bipole to some critical strength to
initiate the buoyancy instability. Convective collapse must have
taken place to enhance the field inside the bipole too.

\begin{acks}
The authors are grateful to the {\it Hinode} team for providing the
data. {\it Hinode} is a Japanese mission developed and launched by
ISAS/JAXA, with NAOJ as a domestic partner and NASA and STFC (UK) as
international partners. It is operated by these agencies in
cooperation with ESA and NSC (Norway). {\it Hinode} SOT/SP
inversions were done at the National Center for Atmospheric Research
(NCAR) under the framework of the Community Spectro-polarimtetric
Analysis Center (CSAC; \url{http://www.csac.hao.ucar.edu/}). This
work is supported by the National Natural Science Foundations of
China (11003024, 10973019, 40974112, 11025315, 11003026, 10873038,
10833007 and 10921303), and the National Basic Research Program of
China (G2011CB811403 and G2011CB811402), and the CAS Project
KJCX2-EW-T07. The authors are indebted to our referee for the very
valuable comments and his kind help to improve the English for the
paper.
\end{acks}

\end{article}

\end{document}